\documentstyle[preprint,prl,aps]{revtex}

\begin{document}

% this places title+abstract over two column widths
%@twocolumnfalse\endcsname  

\draft

\title{A Gauge Invariant Way to Evaluate Quasiparticle Effective Mass 
in Fermionic Systems with Gauge Interactions}

%%Raghav, I propose an alternative title: `A Gauge Invariant Way to
%%Evaluate Quasiparticle Effective Mass in Fermionic Systems with
%%Gauge Interactions'. -- KUN
\author{A. Raghav Chari, F.D.M. Haldane and Kun Yang 
}

\address{Department of Physics, Princeton University, 
Princeton, NJ 08544}

\maketitle

\begin{abstract}

In this paper, we propose a gauge-invariant way to define and calculate 
the effective mass for quasiparticles in systems with gauge interactions, and
apply it to a model closely related to the 
half-filled Landau level problem.
Our model is equivalent to the Halperin-Lee-Read $\nu = 1/2$ Hamiltonian
with an ultraviolet cutoff $\Lambda$ for the gauge fields, and we expand
our answer in powers of $\Lambda / k_F$, assuming it is small.
In this definition the effective mass depends 
only on the gauge-invariant
density and current response functions of the 
system. Within RPA, this definition yields a finite result for the effective 
mass in our model. 
We also comment on corrections to this effective mass formula when 
processes beyond RPA are included.  
Finally, we comment briefly on the 
observation that organizing the perturbation expansion in 
powers of $\Lambda / k_F$ is a way to systematically study the physics 
beyond RPA.
 
\end{abstract}


\section{Introduction}

Recently there has been considerable interest in understanding the novel physics
of a two-dimensional (2D) electron gas in a high magnetic field, with Landau
level filling factor ($\nu$) at or close to one-half.
It was argued\cite{HLR,KZ}
that a useful way to look at the problem is to view electrons
as composite fermions (CF)\cite{jain}
carrying two flux quanta, such that 
the flux carried by the CFs cancel that of the external magnetic field
in average.
Therefore at mean field level the CFs see zero net magnetic flux, and are
expected to form a Fermi-liquid like state, with a Fermi surface satisfying
Luttinger's theorem. Such a simple picture of weakly interacting CFs seems
to agree very well with experiments\cite{willet,goldman,hari,du}, 
at least at a qualitative level,
suggesting that CFs are stable quasiparticles in the system.
Convincing evidence of the existence of a Fermi surface has also been found 
in numerical studies\cite{haldane}. However a more careful analysis\cite{HLR}
(mostly
perturbative) of 
quasiparticle properties found that quasiparticles near the Fermi
surface have a divergent effective mass and very short life time
(except when they interact with super-long range interactions), implying that 
the quasiparticles in this theory are not stable.
Previous studies of quasiparticle properties have been based on the single 
particle Green function or self-energy, which is a {\em gauge dependent}
quantity. It has been shown \cite{KFWL} that singularities encountered
in self-energy calculations disappear in calculations of gauge-invariant
response functions. Another difficulty of the standard perturbative
approach is that the coupling constant for the gauge interaction is not
small in this problem; therefore it is unclear whether or not the 
perturbation series in powers of the coupling contant is convergent. 

In this paper, we propose a
physically motivated definition for the 
effective mass of quasiparticles that is fully gauge-invariant and
expressed in terms of the (gauge-invariant) response 
functions of the system. 
This definition is very much in the spirit of Landau's original 
picture of quasiparticles in the Fermi liquid which arise from adiabatically 
dressing the particles (or holes) in the
free Fermi gas with interactions.
This new definition may be applied to models with gauge interactions 
in general.
In this work 
we apply it to a model that is closely related to the
$\nu=1/2$ problem; the only 
difference is that there is a finite momentum cutoff $\Lambda$
in both the gauge and 
scalar interactions of the CFs. In real space this corresponds to 
a flux that is distributed near the particle within a region of order
$1/\Lambda$ (or a ``fat" flux tube), 
instead of the original point-like distribution of flux. The
original problem is recovered in the limit $\Lambda\rightarrow\infty$, while
in the limit $\Lambda\rightarrow 0$ the mean field approximation
becomes exact. We calculate the leading contribution
to the effective mass from interactions, in the limit of small $\Lambda/k_F$,
which is equivalent to calculating response functions in the Random Phase
Approximation (RPA). We do not find any singularity at this level.
We will also discuss how to go to higher orders in this approach. 
We also suggest that expanding in powers of $\Lambda/k_F$ may be a useful
alternative way to organize the perturbation series.

We start by introducing the model we use, and then present the definition of
the quasiparticle effective mass, and evaluate it at the lowest order in
$\Lambda/k_F$. We conclude with a summary and discussion.

%As a matter of philosophy, we have 
%chosen to sidestep the 
%standard gauge-field treatments of the $\nu = 1/2$ theory 
%because we feel it is not only unnecessary, it also tends to obscure the fact 
%that the only quantities one 
%can safely calculate are gauge-invariant ones (such 
%as response functions); quantities such as the self-energy of the composite 
%fermions are not gauge-invariant, and as such, we prefer not to deal with them.
%% Raghav, I feel the above sentences become unnecessary with the new stuff
%% I added.--KUN
 
\section{Model with Finite Cutoff in Interactions}

Consider the following Hamiltonian for fermions at half-filling :

\begin{equation}
H = \sum_{i} \frac{1}{2m} ( \vec{p_i} - e \vec{A}(\vec{r_i}) - 
\vec{a}(\vec{r_i}))^{2}
+ \frac{1}{2} \sum_{i \neq j} V( \vec{r_i} - \vec{r_j} )
\label{hamiltonian}
\end{equation}

\noindent
where the momenta and positions are for the composite fermions. $\vec{a}$ 
actually depends on the coordinates of the other particles and its form depends 
on the filling factor; it is much easier 
to deal with these quantities 
in momentum space, and we can in fact write (using the explicit expression for 
$\vec{a}$ in terms of the coordinates of the other particles)

\begin{equation}
\delta a^{l} (\vec{r_i}) \equiv e A^{l}(\vec{r_i}) + a^{l}(\vec{r_i}) = 
\sum_{j \neq i} \sum_{q \neq 0} i \Phi (q) \frac{\epsilon^{lm} q^m}{q^2} \exp 
\left(i \vec{q} \cdot (\vec{r_i} - \vec{r_j}) \right)
\label{gaugefield}
\end{equation}

\noindent In the above we have explicitly used the Coulomb gauge for $\vec{a}$.
This is where we introduce the "fat" fluxes: in the exact model, we have
$\Phi (q) = 4 \pi$ for a system at half-filling, but 
in our model we take $\Phi (q) = 4 
\pi$ for $q \leq \Lambda$ and equal to $0$ for $q > \Lambda$. 
This 
corresponds to an ultraviolet cutoff on the gauge field momenta. The
reason for  
doing this will become clear shortly, but for the moment let us remark
that for 
small $\Lambda$ it becomes possible to treat $\Lambda / k_F$ as a small 
parameter in organizing the perturbation expansion of the theory.

We can use equations~\ref{hamiltonian},~\ref{gaugefield} to rewrite the 
Hamiltonian in terms of the density and current operators as follows:

\begin{equation}
H = \sum_{i} \frac{1}{2m} {\vec{p_i}}^{2} + \frac{1}{2} \sum_{q} (\rho (q) 
\rho (-q) - N)V(q) + \left( \frac{1}{2} \sum_{q \neq 0} i \frac{\Phi (q)}{q}  
 \rho (q) (\vec{J}(q) \times \hat{q})^{\dagger} + c.c \right)
\label{ham1}
\end{equation}

Figure~\ref{feynman} shows the Feynman rules for the theory derived from 
equation~\ref{ham1}. For convenience of notation, the interaction lines can 
be encoded into a matrix $\bf{V}$ s.t. 

\begin{equation}
\bf{V} = \left( \matrix{V(q) & -i \Phi (q) /q \cr
i \Phi (q) /q & 0 \cr}\right)
\label{vmatrix}
\end{equation}

In addition, we may also define the following 2-by-2 matrix to simplify the 
writing of the appropriate RPA expressions for the response functions:

\begin{equation}
\bf{K} = \left( \matrix{K_{00} & K_{01} \cr
K_{10} & K_{11} \cr} \right)
\label{kmatrix}
\end{equation}

Here $K_{00}$ is the density-density response function, $K_{01}$ is the 
density-current response function, etc. 
They are physical 
response functions which are independent of the gauge choice of the
Hamiltonian. As a reminder, only the correlation 
functions of the transverse current $\vec{J} (q) \times \hat{q}$ enter this 
matrix, since the longitudinal current is completely determined 
from the density 
(by current 
conservation) and therefore not independent. Note that in the above matrices, 
${\bf K} = {\bf K}(q,\omega)$.

As usual, we may write the RPA response functions ${\bf{K}}^{RPA}$ in terms of 
the free response functions $\bf{K}^{(0)}$ as 

\begin{equation}
{\bf K}^{RPA} = {\bf K}^{(0)} (1 - {\bf V} {\bf K}^{(0)})^{-1},
\label{krpa}
\end{equation}
and the RPA propagator as
\begin{equation}
{\bf V}^{RPA} = ( {\bf V}^{-1} - {\bf K}^{(0)})^{-1}.
\label{rpaprop}
\end{equation}

\noindent
In general, the connected Green functions are obtained from a 
geometric sum over 
the one particle irreducible Green functions (the so-called 1PI Green 
functions), and the RPA response functions are obtained by 
replacing the general 1PI Green functions by the free 1PI contribution. In our 
model with "fat" flux quanta, we can show that the 1PI corrections to the free 
1PI contribution to the response functions are suppressed by factors of $\Lambda 
/ k_F$. This is because the internal interaction lines must have momentum $q 
\leq \Lambda$; this phase space restriction causes the appropriate 1PI gauge 
invariant corrections to the free response functions to be suppressed by 
$\Lambda / 
k_F$. In this sense, when $\Lambda / k_F \rightarrow 0$ the corrections to RPA 
vanish relative the the RPA contributions.

\section{Effective Mass}
Usually, the effective mass of composite fermions is defined in terms of the 
self energy of the composite fermions, which depends on the gauge choice of
the Hamiltonian. Rather than doing this, we would like 
to motivate a very different way to calculate the effective mass. This is based 
on 
an energy theorem discussed by Pines and Nozieres \cite{PN}, and in the
spirit of Landau's original
picture of quasiparticles in the Fermi liquid. If the system is a Fermi
liquid, and its low energy excitations can be decribed as stable quasiparticle
and quasihole excitations, we may label a low energy eigenstate of the 
system by an occupation configuration $n_{\vec{k}}$ (see 
figure~\ref{fermisurface}); such a state is 
smoothly connected to the corresponding state of the free Fermi system, by
{\em adiabatically} turning on the interactions. 
What we
do is to evaluate the energy of such states, which by the Hellman-Feynman 
theorem may be expressed as coupling constant integrals of certain
{\em gauge invariant}
correlation functions. In particular, we calculate 
the difference between the ground state and the excited states with
a single particle-hole pair excitations; by definition this energy 
difference is related to the quasiparticle effective mass.

%What we 
%do is to evaluate the energy of a particular state $\Psi_0$ of the composite 
%fermion system in terms 
%of \emph{gauge-invariant} response functions corresponding to that state. Any 
%state $\Psi_0$, for a fermi liquid, can be defined in momentum space by the 
%momenta of the particles, or in other words, the shape of the fermi surface. 
%Deforming the surface, or creating a quasiparticle-quasihole excitation will 
%change the state  $\Psi_0$ to a new one $\Psi$. 
%We can obtain the effective mass 
%of 
%the composite fermions by calculating the energy difference between the states 
%$\Psi_0$ and the state $\Psi$ (corresponding to a particle-hole excitation). 
%Operationally, this difference will be calculated via a formula for the energy 
%of any state, which is expressed in terms of the response functions of the 
%system in that state.

Let us derive an expression for the energy of a state $\Psi_0$ now \cite{PN}. 
Let $\langle 
\rangle$ denote expectation values with respect to this state. To start with, 
let us rewrite equation~\ref{ham1} by explicitly introducing a parameter 
$\alpha$ which accompanies the coupling constant:

\begin{equation}
H = \sum_{i} \frac{1}{2m} {\vec{p_i}}^{2} + \alpha \left( \frac{1}{2} \sum_{q} 
(\rho (q) \rho (-q) - N)V(q) + \left( \frac{1}{2} \sum_{q \neq 0} i \frac{\Phi 
(q)}{q} \rho (q) (\vec{J}(q) \times \hat{q})^{\dagger} + c.c \right) \right)
\label{ham2}
\end{equation}

\noindent
and therefore we have the energy (as a function of $\alpha$)

\begin{equation}
E_{0}(\alpha)
= \langle \Psi_{0}(\alpha) 
| \sum_{i} \frac{1}{2m}{\vec{p_i}}^2 | \Psi_{0}(\alpha) \rangle 
+ E_{\mathrm{int}}(\alpha)
\end{equation}

\noindent
Differentiating this with respect to $\alpha$ we get

\begin{eqnarray}
\frac{\partial E_0}{\partial \alpha} & = & \langle \Psi_{0} | \frac{\partial 
H}{\partial \alpha} | \Psi_{0} \rangle + \langle \frac{\partial 
\Psi_{0}}{\partial \alpha} | H | \Psi_{0} \rangle + \langle \Psi_{0} | H | 
\frac{\partial \Psi_{0}}{\partial \alpha} \rangle \nonumber \\
& = & \frac{E_{\mathrm{int}}}{\alpha} + E_{0} \frac{\partial}{\partial \alpha} 
\langle 
\Psi_{0} | \Psi_{0} \rangle
\end{eqnarray}

\noindent
Noting that the state is normalized, we get

\begin{equation}
\frac{\partial E_0}{\partial \alpha} = \frac{E_{int}}{\alpha}
\end{equation}

\noindent
This is nothing but a special form of the Hellman-Feynman theorem. 
We may then integrate over $\alpha$ and obtain:

\begin{equation}
E_{0}(\alpha = 1) - E_{0}(\alpha = 0) = \int_{0}^{1} d \alpha 
\frac{E_{\mathrm{int}}}{\alpha}
\label{coupint}
\end{equation}

\noindent
Plugging in equation~\ref{ham2} 

\begin{eqnarray}
\frac{E_{\mathrm{int}}}{\alpha} & = & \frac{1}{2} \sum_{q} V(q)( \langle \rho 
(q) 
\rho^{\dagger}(q) \rangle - N) + \frac{1}{2} \sum_{q \neq 0} \left( \frac{i 
\Phi(q)}{q} \langle \rho(q)(\vec{J}(q) \times \hat{q})^{\dagger} \rangle + c.c. 
\right) \nonumber \\
& = & - \frac{1}{2} \sum_{q \neq 0} V(q) \int_{0}^{\infty} \frac{d \omega}{\pi} 
{\mathrm{Im}} K_{00}(q, \omega) - \frac{1}{2} \sum_{q \neq 0} \int_{0}^{\infty} 
\frac{d \omega}{\pi} \left( \frac{i \Phi(q)}{q} {\mathrm{Im}} K_{01}(q, \omega) 
+ c.c. \right)
\end{eqnarray}

\noindent
Let us note that we may do the $\omega$ integrals via contours (see 
figure~\ref{contours} ; as the figure shows, the contour encloses the real 
excitations of the system):

\begin{equation}
\int_{0}^{\infty} d \omega {\mathrm{Im}}K_{\alpha \beta}(q, \omega) = 
\frac{1}{2i} \oint_{\Gamma} K_{\alpha \beta}(q, \omega) d \omega
\end{equation}

\noindent
We then perform the coupling constant integration (equation~\ref{coupint}) to 
obtain:

\begin{equation}
E_0(\alpha = 1) = E_0(\alpha = 0) - \frac{1}{4 \pi i} \oint_{\Gamma} d 
\omega 
\sum_{q} \int_{0}^{1} d \alpha {\mathrm{Tr}}( {\bf{V}}(q) {\bf{K}}(q,  \omega))
\label{gsenergy}
\end{equation}

\noindent
where we have taken the matrix forms for the interaction and response functions 
from equations~\ref{vmatrix} and~\ref{kmatrix} for compactness of notation.

The response functions $\bf{K}$ depend implicitly on $\alpha$ because 
of interactions. As we mentioned earlier, in the limit $\Lambda / k_F 
\rightarrow 0$ we can replace the full response functions with the corresponding 
RPA expressions. Plugging in from equation~\ref{krpa} the form 
for ${\bf{K}}^{RPA}$ we get:

\begin{eqnarray}
E_{0}(\alpha = 1) & = & E_{0}(\alpha = 0) - \frac{1}{4 \pi i} 
\oint_{\Gamma} d \omega 
\sum_{q} \int_{0}^{1} d \alpha {\mathrm{Tr}} \left( \frac{{\bf{V}}(q) 
{\bf{K}}^{(0)}(q,  \omega)}{1 - \alpha {\bf{V}}(q) {\bf{K}}^{(0)}(q,  \omega)} 
\right) \nonumber \\
& = & E_{0}(\alpha = 0) + \frac{1}{4 \pi i } \oint_{\Gamma} d \omega 
{\mathrm{Tr}} 
\log (1 - {\bf{V}}{\bf{K}}^{(0)} )
\label{gsenergy1}
\end{eqnarray}

\noindent
Of course, $E_0(\alpha = 0)$ is nothing but the energy for free 
fermions (the system being described by the state $\Psi_0$, or alternatively the 
shape of the fermi surface).

We are now in a position to give an expression for the effective mass of 
composite fermions. As we remarked earlier, the energy is a 
functional of the shape of the fermi surface. Explicitly, we have ${\bf{K}} = 
{\bf{K}}[n_k]$ where $n_k$ equals zero for $k$ outside the fermi surface and 
equals one inside. Varying the shape of the fermi surface will vary $\bf{K}$ and 
thereby vary the energy as per equation~\ref{gsenergy1}. 

Let us take the relaxed configuration to be the circular fermi surface, with
$n_{k}^{(0)} = \theta(k_F - |k|)$ and the distorted fermi surface be described 
by 
$n_{k} = n_{k}^{(0)} + \delta n_{k}$ where $\delta n_{k} = \delta_{k,k_{F}+l} - 
\delta_{k,k_{F}-l}$ (i.e, we have created a particle-hole 
pair with particle momentum $k_F + l$ and hole momentum $k_F - l$ : see 
figure~\ref{fermisurface}). For free 
fermions ($\alpha = 0$) the difference between the energies of the two states is 
simply $\delta E_0(\alpha = 0) = 2 k_F \cdot l /m$. It is natural to 
define the effective mass for interacting fermions 
in the same way, i.e, we have 
\begin{equation}
\delta E_0(\alpha = 1) = \frac{2 k_F \cdot l}{m^{*}}
\end{equation}

\noindent
We like this definition of the effective mass because it is defined in terms of 
gauge-invariant quantities. Let us also point out that in the sums over $q$ in 
the above expressions there is a cutoff $\Lambda$ in our model, because $\Phi(q) 
= 0$ for $q > \Lambda$; in addition, we require $V(q) = 0$ for $q 
> \Lambda$ as well. Then we get

\begin{equation}
\frac{2 k_F \cdot l}{m^{*}} = \frac{2 k_F \cdot l}{m} - \frac{1}{4 \pi} 
\sum_{q \leq \Lambda} \oint_{\Gamma} d \omega \frac{V(q) \delta K_{00} 
+ \frac{\Phi^{2}}{q^2}(K_{00} \delta K_{11} + \delta K_{00} K_{11})}{1 - 
V(q)K_{00} - \frac{\Phi^{2}}{q^2} K_{00}K_{11}}
\end{equation}

\noindent
where $K_{\alpha \beta}[n_k] = K_{\alpha \beta}[n_k](q, i \omega)$ and $\delta 
K_{\alpha \beta} = K_{\alpha \beta}[n_k] - K_{\alpha \beta}[n_{k}^{(0)}]$ 
expanded to linear order. We can then evaluate the above expressions in the 
small $\Lambda$ limit. As is clear, while doing the $\omega$ integral, we will 
pick up (in general) contributions from all the poles in the integrand; however, 
in the small $q$ limit, the cyclotron mode saturates the F-sum rule, and 
therefore this is the only pole we need to take into account in the limit 
$\Lambda \rightarrow 0$. 
First, we evaluate the effective mass as a function of $\Lambda$ when $V = 0$. 
Then we write down the correction to the effective mass treating $V$ as a small 
perturbation. The calculations are straightforward, and we get for $V = 0$:

\begin{equation}
\frac{1}{m^{*}} = \frac{1}{m} \left( 1 + \frac{\Lambda^{4}}{2k_F^{4}} \right)
\label{mstar}
\end{equation}

\noindent
For small $V$ we will get 

\begin{equation}
\frac{1}{m^{*}} = \frac{1}{m} \left( 1 + \frac{\Lambda^{4}}{2k_F^{4}} - 
\frac{3}{8 \pi} \int_{0}^{\Lambda} \frac{mV(q) q^{5}}{k_{F}^{6}} dq \right)
\label{effectivemass}
\end{equation}

We should emphasize that there is no divergence in the effective mass at this 
level. Also, we should clarify that the $i \omega \sim q^3$ pole in the RPA 
interaction (found for $q$ small and $\omega \ll q v_F$) does not contribute to 
our effective mass formula. This is easily seen because the energy 
depends only on physical excitations of the system, given by poles of the 
response functions that lie on the real axis, and the contour $\Gamma$ is chosen 
to only include such physical excitations. 
Corrections to RPA could certainly, 
in principle, suffer a divergence due to this pole (when we have to do 
additional frequency and momentum integrals for internal interaction lines). 
However, we found no divergences in the 
physical response functions when we included the 2-loop self-consistent 
corrections to RPA for the 
response functions. We remark once again, that in our approach, corrections to 
the effective mass formula from such processes will be higher order in our 
expansion in $\Lambda / k_F$. 

To conclude this section, let us very briefly 
point out the relation between our 
definition of the effective mass versus the standard definition coming from the 
self energy of the composite fermion. The exact theory, {\em a la}
Halperin, Lee and 
Read \cite{HLR} has interacting fermions 
and gauge fields. There are two ways to evaluate the free energy of the system. 
One can first integrate out the gauge fields and obtain an effective action for 
the \emph{fermions}, from which one then evaluates the free energy for system in 
the saddle point approximation. This process basically involves calculating the 
self energy of the composite fermion, and seeing how this self-energy changes 
the free energy of the system of composite fermions. This gives one definition 
of the effective mass, which is the one we are used to. Alternately, we may 
integrate out the fermions first, and obtain an effective action for the 
\emph{gauge fields}, and again one can find the free energy of this system in 
the saddle point approximation. This would give us a formula for the free energy 
that is identical to the expression given in equation~\ref{gsenergy1}. If one 
integrated out the fermi and gauge fields exactly, one would get the same 
answer. However, because we are making the saddle point approximation in one 
case with the effective fermion action and in the other case with the effective 
gauge field action, there is no guarantee that we will get the same result. The 
fact that in the latter approach (the one we use) we only make reference to 
gauge invariant finite quantities makes it somewhat more desirable and possibly 
more physically meaningful.

\section{Discussion}
In the above, we have described a different way to define the effective mass of 
quasiparticles in the system, first described by Pines and Nozieres in their 
classic book\cite{PN}, in the context of degenerate electron gas. 
This method of calculating the effective mass is 
especially suitable for systems with gauge degrees of freedom, because things 
are defined purely in terms of gauge invariant observables.

We should comment that 
our expression~\ref{effectivemass} seems to indicate that 
for increasing $\Lambda$ (and $V = 0$) the effective mass decreases 
monotonically. Of course, 
this is only for small $\Lambda$. In general, we have fermions in a strong 
magnetic field (and more specifically, fermions in the lowest Landau level) and 
as Shankar and Murthy \cite{SM} suggest, it should be possible to see signs of 
this even if the Chern-Simons formulation reduces the theory to quasiparticles 
in weak magnetic field. For fermions in the lowest Landau level, however, the 
kinetic energy is quenched; initially, 
we were hoping that our calculation would 
show signatures of this quenching by a monotonic enhancement of the effective 
mass as a function of 
increasing $\Lambda$ (in the lowest Landau level, $m^{*} = 
\infty$ in the absence of interactions), but we find exactly the opposite to 
lowest order. Still, for larger $\Lambda \sim {\cal{O}}(k_F)$ we would have to 
include contributions to the effective mass beyond RPA, and it is possible that 
the effective mass could show a monotonic enhancement of the desired kind.

Lastly, let us remark that $\nu = 1/2$ model with "fat" flux quanta has the 
merit that we can treat perturbation theory in a systematic fashion despite the 
fact that the coupling constant in the theory is not small. In a loose sense, 
adding interaction lines to 1PI diagrams results in a phase space suppression 
$\Lambda / k_F$ for each internal 
interaction line (by interaction, we mean both 
$\rho \rho$ as well as $\rho J$). This is because the momentum of each such 
internal line is restricted to be smaller than $\Lambda$. We have explicitly 
checked this result for all 2-loop diagrams (using the RPA interaction instead 
of the bare interaction, in order to treat the problem self-consistently). 
Whether or not the results obtained using our small $\Lambda$ approximation are 
meaningful for large $\Lambda$ (the exact 
problem has $\Lambda = \infty$) 
is an issue we cannot address within the present framework. However, the 
\emph{method} of defining the effective mass in systems with gauge interactions 
in terms of gauge-invariant response functions as above is certainly 
well-defined and completely general.

This work is supported by NSF grant DMR-9400362.

\begin{figure}
\caption{\label{feynman}
Feynman rules for theory.}
\end{figure}

\begin{figure}
\caption{\label{fermisurface}
On the left, the ground state of the system is shown, for which we have a
circular fermi surface. On the right, we depict a quasiparticle-quasihole
excitation of this state. The text describes how we calculate the energy
difference between these states in our model, and the effective mass is
defined in terms of this energy difference.}
\end{figure}

\begin{figure}
\caption{\label{contours}
The contour includes all the physical modes of the system. As explained in the
text, poles in the response function may appear off the real axis, but these
will not contribute to the integral.}
\end{figure}

\end{document}